\pdfoutput=1
\documentclass[pss]{wiley2sp} 
\usepackage{amsmath}

\begin{document}

\title{Circular dichroism in angle-resolved photoemission spectroscopy of topological insulators}

\titlerunning{CD-ARPES of TI}

\author{%
    Yihua Wang and Nuh Gedik\textsuperscript{\Ast}
    }
\authorrunning{Y.\,H. Wang and N. Gedik}

\mail{e-mail
  \textsf{gedik@mit.edu}, Phone:
  +617-253-3420, Fax: +617-258-6883}

\institute{%
Department of Physics, Massachusetts Institute of
Technology, Cambridge MA 02139, USA
}
\received{XXXX, revised XXXX, accepted XXXX} 
\published{XXXX} 


\abstract{%
%
%
%
\abstcol{%
  Topological insulators are a new phase of matter that exhibits exotic surface electronic properties. Determining the spin texture of this class of material is of paramount importance for both fundamental understanding of its topological order and future spin-based applications. In this article, we review the recent experimental and theoretical studies on the differential coupling of left- versus right-circularly polarized light to the topological surface states in angle-resolved photoemission spectroscopy. These studies have shown that the polarization of light and the experimental geometry plays a very important role in both photocurrent intensity and spin polarization of photoelectrons emitted from the topological surface states. A general photoemission matrix element calculation with spin-orbit coupling can quantitatively explain the observations and is also applicable to topologically trivial systems. These experimental and theoretical investigations suggest that optical excitation with circularly polarized light is a promising route towards mapping the spin-orbit texture and manipulating the spin orientation in topological and other spin-orbit coupled materials.
  }
  }

\titlefigure[height=7cm,width=6cm]{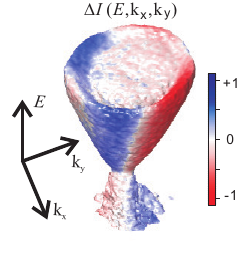}
\titlefigurecaption{%
  The circular-dichroic angle-resolved photoemission spectrum of Bi$_2$Se$_3$ as obtained by the difference between spectra taken with left- vs. right- circularly polarized light using a time-of-flight electron spectrometer.}

\maketitle   

\section{Introduction}
\subsection{Topological Insulators}
Three-dimensional topological insulators (3D TI) are a new phase of matter with a bulk band gap but a conductive surface \cite{MooreReview,KaneReview,QiZhangReview}. The nontrivial topology of its bulk wavefunction, as defined by a topological invariant, guarantees the surface states to cross the Fermi level odd number of times between two time-reversal invariant momenta. Such topological protection leads to a linear energy-momentum dispersion of the surface states similar to that of the massless Dirac fermions. Such surface band structure in the shape of a Dirac cone may allow many future opto-electronic applications of TI \cite{Hosur,HsiehSHG,McIver,YHWang12}.

Distinct from their two-dimensional version, 3D TI allows direct determination of their topological invariant by measuring their bandstructure with surface-sensitive scattering probes such as angle-resolved photoemission spectroscopy (ARPES) \cite{Hsieh08,Chen,Xia}. ARPES is a photon-in electron-out technique which maps the electronic band structure of a material by measuring the kinetic energy and momentum of the photoemitted electrons \cite{Hufner}. The first discovered 3D TI Bi$_{1-x}$Sb$_x$ has surface bands crossing the Fermi level five times and a rather small band gap \cite{Hsieh08}. Bi$_2$Te$_3$ \cite{Chen} and Bi$_2$Se$_3$ \cite{Xia,ZhangNatPhy,HsiehNat09} were later found to possess a single surface Dirac cone. Because of their simple surface electronic structure and large bulk band gap, which manifests the topological behavior even at room temperature, Bi$_2$Te$_3$ and Bi$_2$Se$_3$ have become the reference materials for the current studies on 3D TI and will be the focus material of this article.

Besides their unique band structure, the surface electrons of TI are spin-polarized because of the strong spin-orbit interaction in the bulk. The broken inversion symmetry at the surface dictates the spin-splitting of the surface states \cite{Bychkov}. The spin polarization is locked to their momentum perpendicularly, forming a chiral spin texture in the momentum space with the sense of chirality reversed across the Dirac point (DP). Surface electrons are also immune from back scattering off disorder and non-magnetic impurities as a result of their unique spin texture. The spin-polarized surface electrons of TI provide a promising platform for many applications ranging from dissipationless spin-based electronic devices (spintronics) \cite{Wolf} to quantum computation \cite{KaneReview,QiZhangReview}.

\subsection{Spin polarization of topological surface states measured by spin-resolved ARPES}
The spin-momentum locking of the surface states of TI was first demonstrated using spin-resolved ARPES \cite{HsiehNat09}. Hsieh et al. have shown that the surface electrons in Bi$_2$Se$_3$ with opposite momentum have in-plane anti-parallel spin polarizations, which are perpendicular to the momentum \cite{HsiehNat09}. Subsequent spin-resolved ARPES experiment by Souma et al. \cite{Souma} shows that a spin component normal to the surface plane is present in Bi$_2$Te$_3$.

In a spin-resolved ARPES experiment, the spin polarization of the photo-electrons collected by the electron spectrometer are determined by their scattering angle with a gold foil \cite{Osterwalder}. Spin-resolved ARPES has played a crucial role in studying the electronic and spin structure of Rashba type spin-splitting in both topological and non-topological materials \cite{Meier}. One challenge with this technique is its intrinsically low efficiency because of the small spin-dependent scattering cross section. Typically measurement along only one of the dimensions of the energy-momentum space is acquired within the sample lifetime.

Aside from the challenge in efficient data acquisition, spin-resolved ARPES experiments have not been able to determine the exact spin-polarization of the surface states. A first-principle calculation shows that the surface states are 50$\%$ spin-polarized and not energy dependent in both Bi$_2$Se$_3$ and Bi$_2$Te$_3$ \cite{Yazyev}. However, different spin-resolved ARPES measurements have measured spin polarization of surface states ranging from 20$\%$ \cite{HsiehNat09} to more than 85$\%$ on Bi$_2$Se$_3$ \cite{Souma,Pan,Xu,Jozwiak}. While lower experimental values are possible due to unpolarized background and finite resolution of the apparatus, fully spin polarized surface states is not expected in a spin-orbit coupled system. This is because with the presence of spin-orbit coupling neither electron spin nor orbital angular momentum is a good quantum number and total angular momentum (pseudospin) has to be employed in describing the wavefunctions \cite{Sakurai}. Nevertheless, the expectation value of spin in the topological surface states is directly proportional to its pseudospin \cite{FuHexWarp,KaneMele,ParkLouie} and for that reason they are often interchangeably used in TI literature.

\subsection{Circular dichroic photocurrent and optical spin orientation}
Because a photoemission process is an optical transition, both the photocurrent intensity and the final state spin polarization depends on the polarization of light \cite{OpticalOrientation}. Photon helicity dependent photoemission intensity has long been used to probe the symmetry of the initial state wavefunction in various systems including molecules \cite{Westphal}, heavy fermion materials \cite{Vyalikh}, high temperature superconductors \cite{Kaminski,Borisenko,Zabolotnyy} and recently graphene-related materials \cite{Liu}. In spin-orbit coupled systems, the helicity of light can couple to the spin degree of freedom through the total angular momentum \cite{OpticalOrientation,Schneider}. Excited by left- versus right-circularly polarized light, the photocurrent from magnetic materials shows magneto-dichroic intensity \cite{Magnetism,Schneider}, which has been widely used in microscopy \cite{Stoher} and spectroscopy \cite{Schneider91} of these materials.

Besides the photocurrent intensity, the spin-polarization of the photoelectrons also depends on the helicity of light. For non-magnetic metals with spin-orbit coupling, photoelectrons with non-zero spin polarization are measured by spin-resolved ARPES \cite{Schneider,Halilov,Frentzen}. Circularly polarized light can also populate conduction band of semiconductors with spin polarized electrons and is in the essence of recent developments in optical spin orientation \cite{MoorePhotoCurrent,Wolf,OpticalOrientation}.

\begin{figure}[tb]
\includegraphics[scale=0.18]{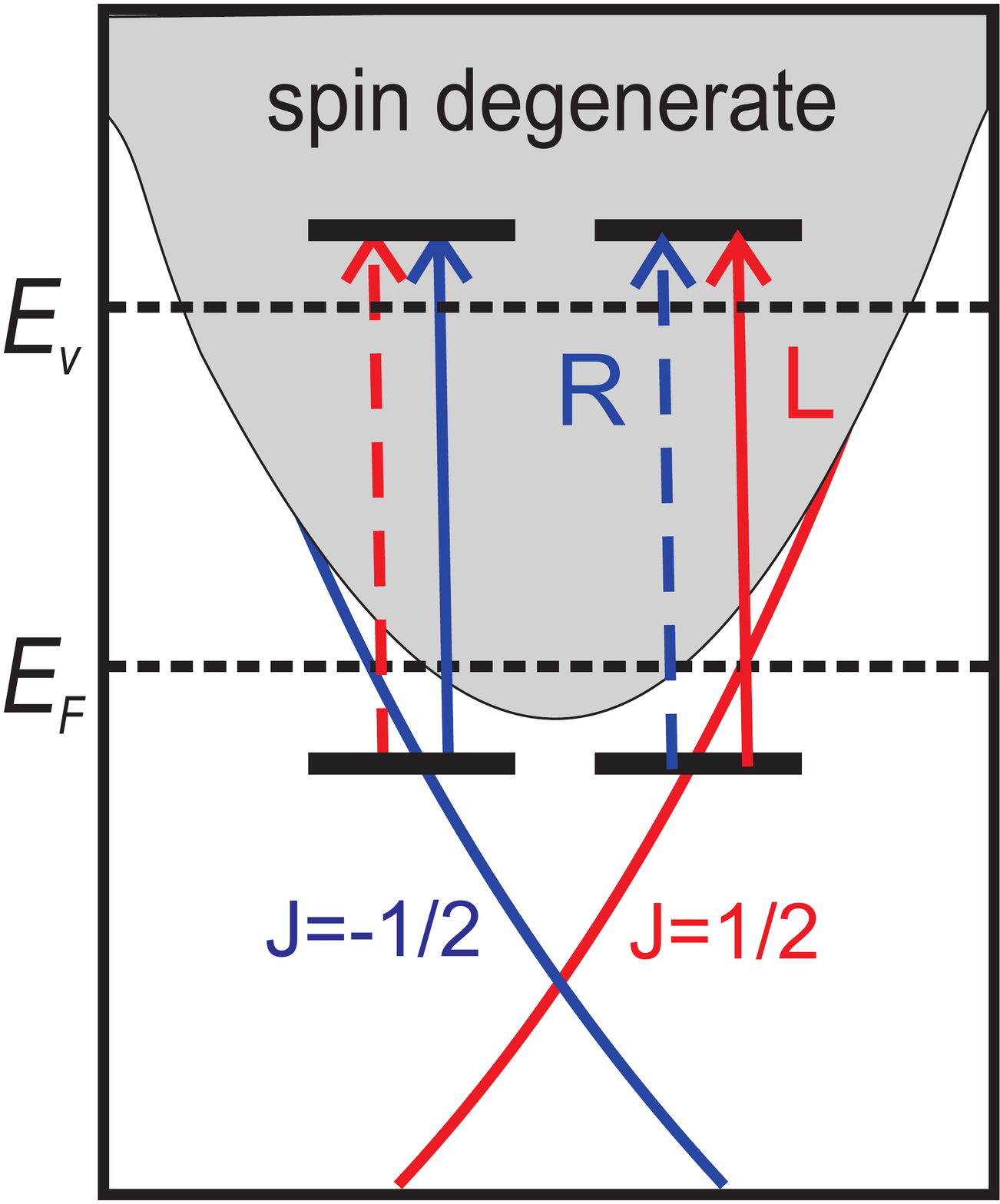}
\caption{ARPES optical transition diagram with circularly polarized light. The spin polarization of the initial state and the helicity of light determines the matrix element and the intensity (shown as solid and dashed lines).}
\label{fig:CDOpticalTransition}
\end{figure}

\begin{figure}[htb]
\includegraphics[width=\columnwidth]{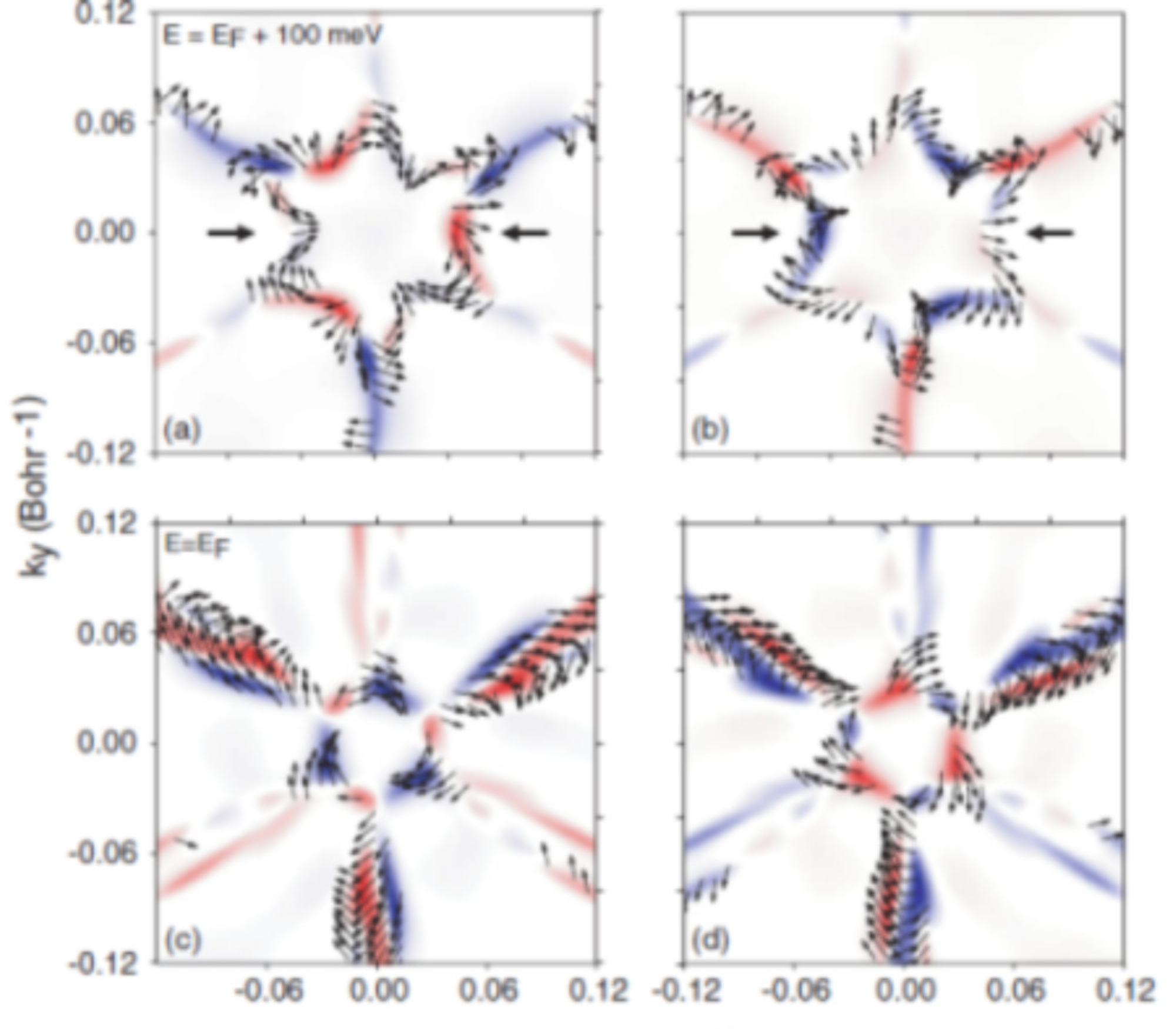}
\caption{Spin polarization of photoelectrons in dichroic photoemission from Bi$_2$Te$_3$ based on first principle calculations. Left (right) column: LCP (RCP) light. Top (bottom) row: $E_F$ ($E_F$ + 100 meV). The color scale represents the z component of the spin difference of the photoelectron, with (red, white, blue) = (negative, zero, positive) values. The in-plane xy components are visualized as small arrows. Bold arrows mark the arcs of the warped constant energy contours (top row) that are discussed in the text. From Ref. \cite{Mirhosseini}.} \label{fig:MirhosseiniFig2}
\end{figure}

Applying the principle of optical spin orientation to topological insulators, which also have strong spin-orbit coupling, we can see that spin polarization of photoelectrons emitted from the surface states will similarly depend on the light polarization and experimental geometry. By calculating the photoemission matrix element with a light-coupled Hamiltonian of a spin-orbit system [Fig. \ref{fig:CDOpticalTransition}], first principle calculations have shown that spin polarization between the crystal electrons and photoemitted electrons can be different \cite{Mirhosseini,ParkLouie}. In fact, 100$\%$ spin-polarized electrons can be emitted from the topological surface states of TI if polarized photons are employed even though the topological surface states are not fully spin-polarized \cite{ParkLouie}, which is a direct consequence of the nondegeneracy of topological surface state. The aforementioned discrepancies in the magnitude of spin polarization of photoemitted electrons from topological surface states measured by spin-resolved ARPES \cite{Souma,Pan,Xu,Jozwiak} are likely due to the optical spin orientation effect caused by differences in light polarization and setup geometry. Furthermore, Fig. \ref{fig:MirhosseiniFig2} shows that not only the magnitude of the spin polarization of the photoelectrons but also the direction of it depends on the helicity of light \cite{ParkLouie,Mirhosseini}.

Knowing that the photoelectron spin-polarization may be different from that of the initial state, the circular dichroic ARPES spectra (CD-ARPES) may be an alternative for probing the spin-orbit texture of topological surface states. This will be shown later from a matrix element calculation of the photoemission process, where the helicity-dependent intensity directly relates to the spin polarization of surface states.

\section{Mapping of topological spin-orbit texture with CD-ARPES}
\subsection{CD-ARPES from the surface states of TI}

\begin{figure}[htb]
\includegraphics[scale=0.3]{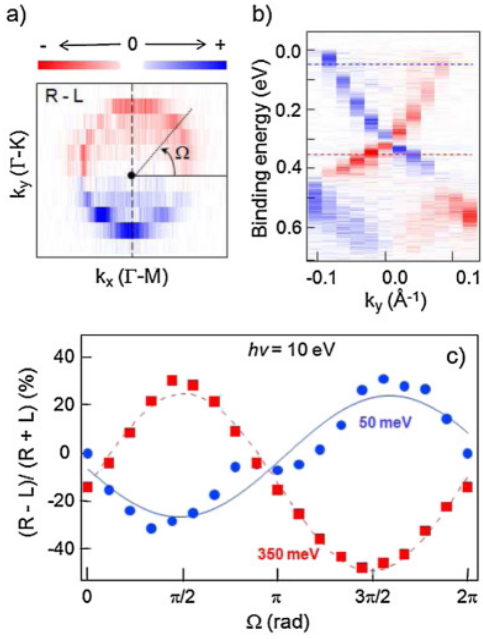}
\caption{Bi$_2$Se$_3$ ARPES difference between spectra taken with right- and left-circularly polarized light. (a) shows a constant energy map close to the Fermi level. (b) is an energy-momentum spectra along $\Gamma$-K direction. (c) is the radially integrated constant energy difference spectra shown in (a) as a function of angle $\Omega$ along the Fermi contour. From Ref. \cite{Park2012}.} \label{fig:Park}
\end{figure}

\begin{figure}[htb]
\includegraphics[width=\columnwidth]{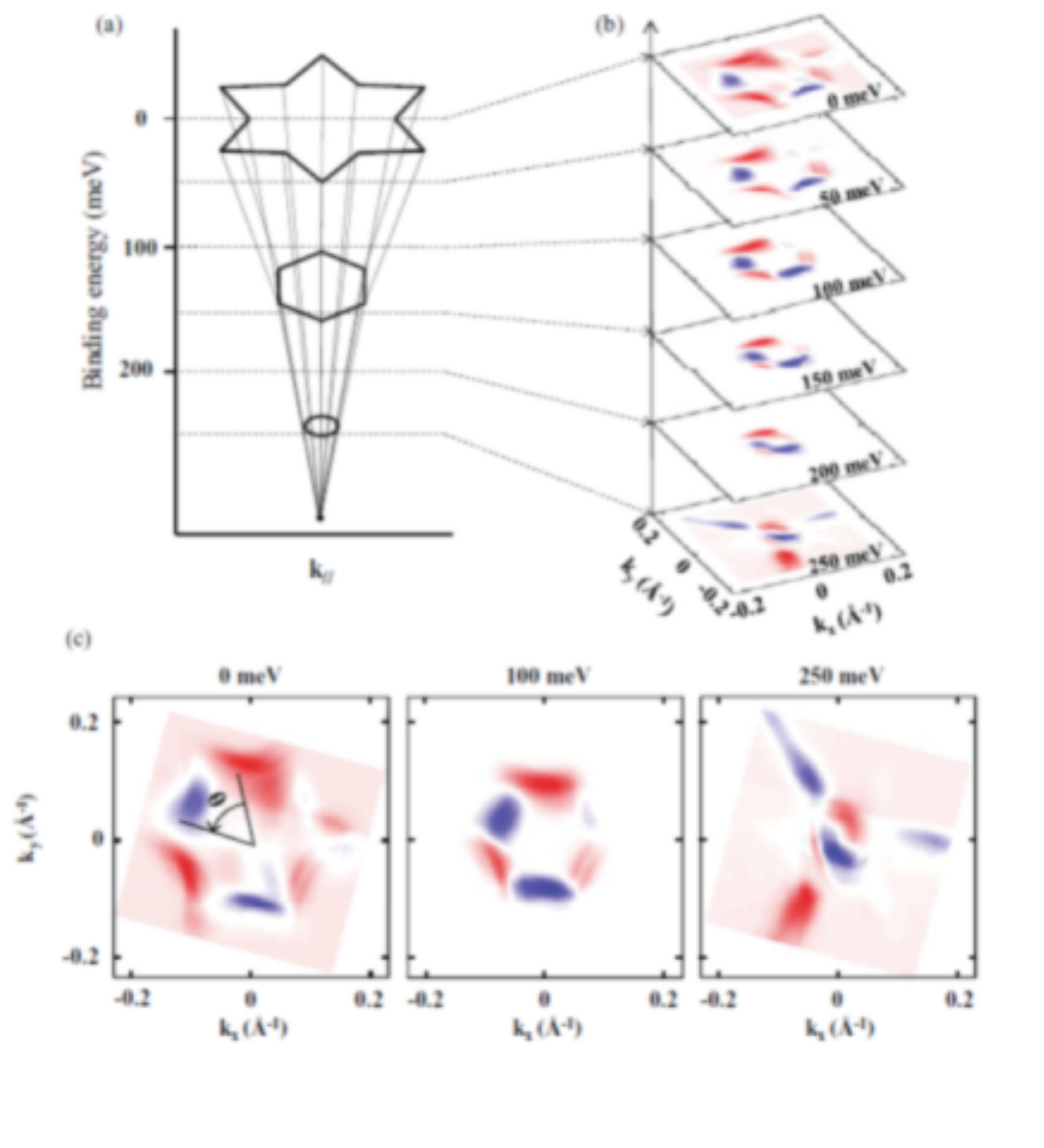}
\caption{Circular dichroism ARPES difference spectra from Bi$_2$Te$_3$. (a) Illustration of Dirac cone of Bi$_2$Te$_3$. (b) Dichroism in the momentum space at various binding energies. (c) Constant dichroism map at selected binding energies of 0, 100, and 250 meV. From Ref. \cite{Jung}.} \label{fig:Jung}
\end{figure}

ARPES provides the capability to directly identify the states which exhibits circular dichroic intensity. A few recent ARPES studies on topological insulators using circularly polarized photons observe strong CD from the surface states but very weak bulk contribution \cite{Jung,Park2012,Ishida,Scholz,YHWang}. These studies focus on the prototypical materials Bi$_2$Se$_3$ and Bi$_2$Te$_3$, where the surface states have a simple bandstructure composed of a single Dirac cone \cite{KaneReview}. As can be seen from the difference spectra $\Delta I (E,k)$ obtained by subtracting ARPES spectra taken with left ($I_L$) vs. right ($I_R$) helicity [Figs. \ref{fig:Park}(b),\ref{fig:Jung}(b),\ref{fig:YHWFig1}(d). (Also Fig. 3(a) in Ref. \cite{Scholz} and Fig. 3(b) in Ref. \cite{Ishida})], the surface Dirac cone consistently exhibit momentum-dependent CD photoemission intensity with the sign switched across the DP. The bulk states, which can be clearly seen in the ARPES spectra [Fig. \ref{fig:YHWFig1}(b) (Also Fig. 1 in Ref. \cite{Jung})], is absent in the difference spectra. It is worth noting that such features persist even though the photon energy used in these studies varies from soft x-ray of the synchrotron radiation source \cite{Jung,Park2012,Ishida,Scholz} to the deep ultra-violet regime of laser source \cite{YHWang}. This strongly suggests that the circularly dichroic ARPES spectra of topological insulators mainly originate from the initial states of the optical transition in the photoemission process.

Sharing this general feature, the CD-ARPES spectra from the surface states of Bi$_2$Se$_3$ and those of Bi$_2$Te$_3$ differ in details \cite{Park2012,Ishida,Scholz,YHWang}. Park et al. \cite{Park2012} have observed CD patterns on Bi$_2$Se$_3$ which reverses sign across the photon incident plane [Fig. \ref{fig:Park}(a)]. Furthermore, the radially integrated difference spectra follow a sinusoidal dependence on the angle around the constant energy contour [Fig. \ref{fig:Park}(c)]. In the case of  Bi$_2$Te$_3$, Jung et al. \cite{Jung} have observed a component with $\sin(3\theta)$ angular dependence in the constant energy difference intensity maps in addition to the $\sin(\theta)$ component [Fig. \ref{fig:Jung}]. The much stronger warping effect \cite{FuHexWarp} in Bi$_2$Te$_3$ than in Bi$_2$Se$_3$, as can be seen from the shape of their constant energy contours \cite{Chen,Kuroda}, suggests that the modulations in the CD-ARPES spectra of Bi$_2$Te$_3$ is a result of hexagonal warping \cite{Jung}.

\begin{figure}[htb]
\includegraphics[width=\columnwidth]{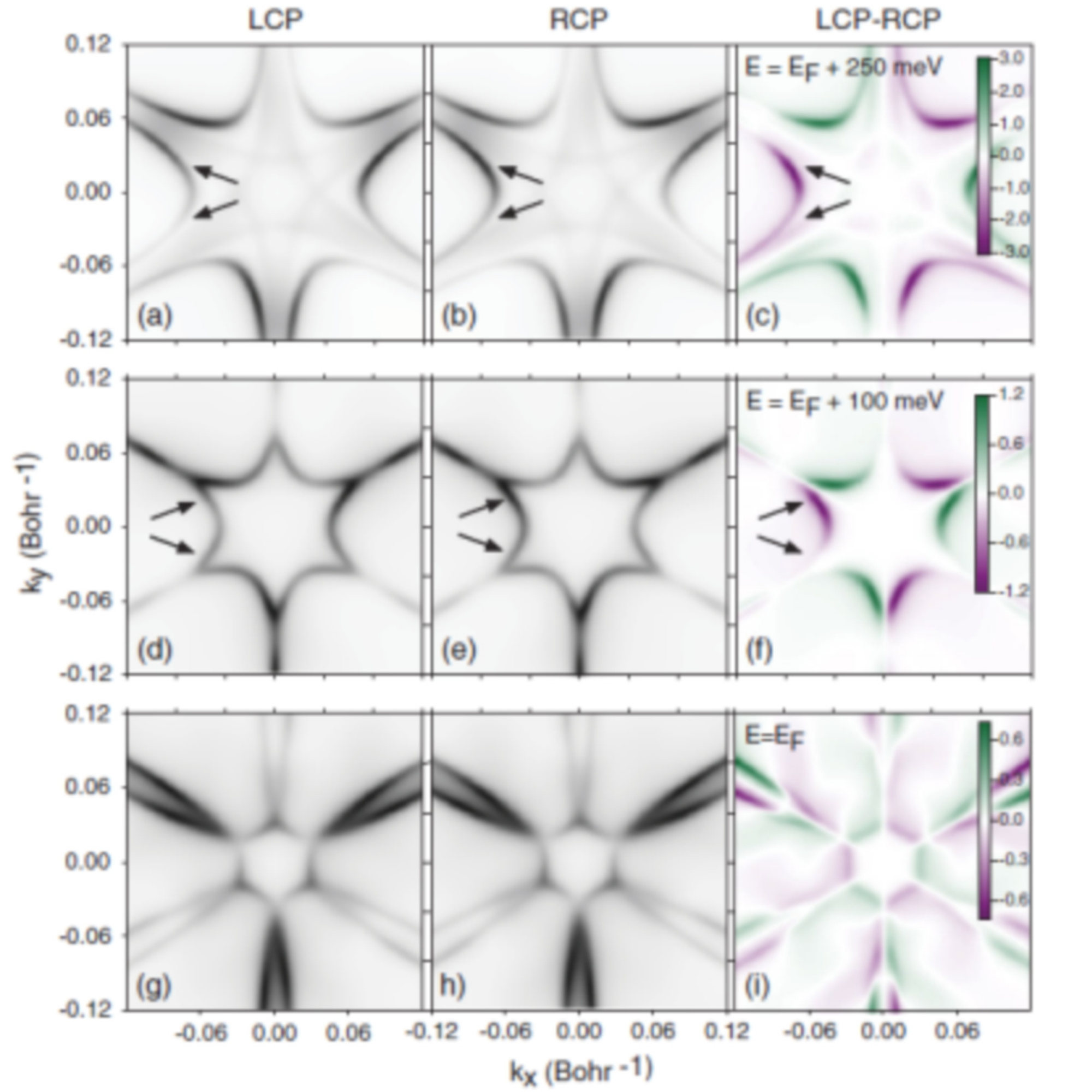}
\caption{Circular dichroism from Bi$_2$Te$_3$ by first-principle calculations. Left and center column: Intensities for left and right circularly polarized light, respectively, shown as gray scale. Right column: circular dichroic intensity difference, represented as color scale. Data for each energy share a common intensity scale. Arrows mark intensity asymmetries within selected arcs of the warped constant energy contours (a)–(f). From Ref. \cite{Mirhosseini}.} \label{fig:MirhosseiniFig1}
\end{figure}

\subsection{Geometric effect on CD-ARPES spectra}

The exact geometry of the experimental setup is very important for direct comparison with theory \cite{Hufner}. Ishida et al. have found that the CD pattern in Cu-doped Bi$_2$Se$_3$ is very similar to a strongly correlated material SrTiO$_3$ \cite{Ishida}. Part of the similarity can be attributed to the identical experimental geometry at which both measurements have been performed (see Fig. 1 in Ref. \cite{Ishida}). By keeping the same geometry as the measurement on Bi$_2$Te$_3$ by Jung et al. \cite{Jung}, the first principle calculation by Mirhosseini et al. \cite{Mirhosseini} qualitatively reproduces the CD-ARPES spectra which deviates from the ideal sinusoidal form [Fig. \ref{fig:Jung}], as can be seen in Fig. \ref{fig:MirhosseiniFig1}.

\begin{figure}[tb]
\includegraphics[width=\columnwidth]{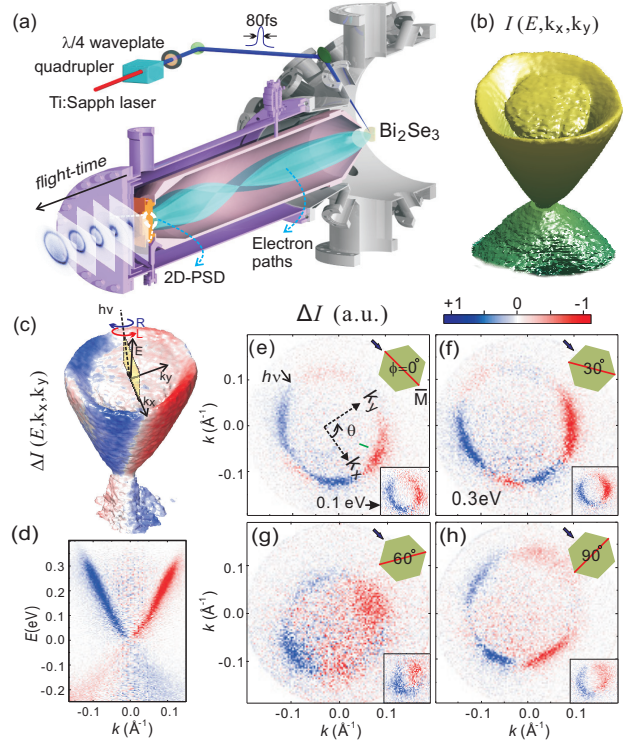}
\caption{CD-ARPES spectra of Bi$_{2}$Se$_{3}$ obtained by a TOF spectrometer and laser source. (a) Schematic of a time-of-flight based angle-resolved
photoemission spectrometer. PSD: position-sensitive detector. (b) A
typical iso-intensity surface in $(E,k_x,k_y)$ space from
Bi$_2$Se$_3$ collected simultaneously using linearly polarized
photons. (c) Difference of TOF-ARPES data measured using right- and
left-circularly polarized light. $h\nu$ denotes the incident photon
direction. (d) $E-k$ cut through the CD data volume in (c) along the
angle denoted by the green dash in (e). (e-h) Constant energy slices
through (c) at the Fermi level and at 0.1eV above the DP (insets)
for sample rotation angles of $\phi=0^{\circ}, 30^{\circ},
60^{\circ}$ and $90^{\circ}$ respectively. Green hexagons represent
the Brillouin zone of Bi$_{2}$Se$_{3}$(111), red lines are the
mirror planes and the arrow denotes the photon incident direction. From Ref. \cite{YHWang}.
}
\label{fig:YHWFig1}
\end{figure}

In order to systematically study how the CD-ARPES spectra vary with geometry, Ref. \cite{YHWang} has performed CD-ARPES on Bi$_2$Se$_3$ at various geometries with a time-of-flight (TOF) electron spectrometer. Unlike a hemispherical electron energy analyzer \cite{Hufner}, a TOF spectrometer [Fig. \ref{fig:YHWFig1}(a)] allows simultaneous measurement of energy $E$ and both in-plane momentum $k_x$ and $k_y$ of the photoelectrons without sample or detector rotation \cite{Kirchmann,Kromker,Carpene,Winkelman}, as shown in Fig. \ref{fig:YHWFig1}(b). The energy of the electrons are measured by the flight-time through the flight-tube and the momenta are imaged by the 2D position-sensitive detector Fig. \ref{fig:YHWFig1}(b). The capability of keeping the same sample-light orientation allows direct comparison of the initial states without having to compensate for the polarization change while electrons with different momentum are measured.

The CD spectrum taken with a TOF spectrometer at various geometries is shown in Fig. \ref{fig:YHWFig1}. Consistent with other results, the bulk states do not exhibit CD regardless of the sample rotation [Figs. \ref{fig:YHWFig1}(e)-(h)]. In agreement with the symmetry requirement, the strong CD on the surface states is odd about the photon incidence plane when it is aligned with the mirror plane of the sample [Figs. \ref{fig:YHWFig1}(c) and (e)]. The difference also switches sign across the DP [Figs. \ref{fig:YHWFig1}(c) and (d)]. At energies far from the DP [Figs. \ref{fig:YHWFig1}(c) and (e)], there are certain modulations in the difference spectra, which is not present in energies close to the DP [Figs. \ref{fig:YHWFig1}(e) inset]. This modulation at higher energies changes as a function of sample rotation angle $\phi$ [Figs. \ref{fig:YHWFig1}(e)-(h)] whereas at energies close to the Dirac point the difference spectra are independent of $\phi$ [Figs. \ref{fig:YHWFig1}(e)-(h) insets]. The CD-ARPES feature at low energies independent of sample rotations suggests that it has the same origin as the common feature shared among experiments with different geometries and materials [Figs. \ref{fig:Park}(b),\ref{fig:Jung}(b),\ref{fig:YHWFig1}(d)], which is closely related to the ideal spin texture of the topological surface states.

The geometry-dependent features in the CD-ARPES spectra of Bi$_2$Se$_3$ [Figs. \ref{fig:YHWFig1}(e)-(h) insets] are reminiscent of the out-of-plane spin texture measured by spin-resolved ARPES \cite{Souma}. Scholz et al. have simultaneously performed CD-ARPES and spin-resolved ARPES on both Bi$_2$Se$_3$ and Bi$_2$Te$_3$ \cite{Scholz}. From spin-resolved ARPES, they have found that the photoemitted electrons can be fully polarized, consistent with some of the earlier results \cite{Souma,Pan,Xu,Jozwiak}. By comparing their CD-ARPES spectra with their spin-resolved measurements, they show that the two have similar energy dependence. This finding suggests that circularly polarized light can couple to the spin of topological surface states through spin-orbit interaction.

\subsection{Photoemission matrix element calculation of a spin-orbit coupled system}

Through the inclusion of a vector potential and spin-orbit coupling in the matrix element calculation, a quantitative relationship between the initial state and circular dichroic difference intensity can be clearly shown \cite{YHWang,ParkLouie,Bian}. The Hamiltonian for a system with spin-orbit coupling is given by \cite{Dresselhaus}:
\begin{equation}
H= \frac{{\vec P}^2}{2m} + V(\vec{r}) + \frac{\hbar}{4m^2 c^2}
({\vec P} \times \vec \nabla V) \cdot  \vec s
\end{equation}
where $\vec P$ is momentum operator, $V(\vec{r})$ is the crystal
potential, and $\vec s$ is the electron spin. Coupling to an
electromagnetic field is obtained via ${\vec P} \rightarrow {\vec P}
- e {\vec A}$, where $\vec{A}$ is the photon vector potential, such
that to first order in $\vec{A}$:
\begin{equation}
H({\vec A}) = H - \vec{\cal P} \cdot {\vec A} \label{ha}
\end{equation}
where $\vec{\cal P}\equiv\frac{e}{m} {\vec P} - \frac{\hbar e}{4m^2
c^2} ({\vec \nabla} V \times {\vec s})$. The photoemission matrix
element between the initial and final states is given by
\begin{equation} \label{m}
M(\vec{k},f) = \langle f_{\vec k} | \; \vec{\cal P} \cdot \vec{\cal
A}|\vec{k}\rangle
\end{equation}
where $\vec{\cal{A}} \equiv \int d t {\vec A}(t) e^{i\omega t}$ is
the Fourier transform of $\vec{A}$ and $|f_{\vec k}\rangle$ is the
spin-degenerate final state. The initial state $|\vec{k}\rangle=u_{\vec{k}} |
\phi^i_+ \rangle  + v_{\vec{k}} | \phi^i_-\rangle$ is a linear
combination of two-fold degenerate pseudospin states $|\phi^i_\pm
\rangle$ at the DP that are eigenstates of pseudospin. Such description of the surface states is widely used in standard $k\cdot p$
theory \cite{KaneReview,ZhangNatPhy}.The coefficients $u_{\vec{k}}$ and $v_{\vec{k}}$
determine the expectation value of three pseudospin components: $
\langle S_x \rangle_{\vec k}= \hbar(u_{\vec k}^* v_{\vec k}
+v_{\vec k}^* u_{\vec k})$, $\langle S_y\rangle_{\vec
k}=\hbar(v_{\vec k}^* u_{\vec k} - i  u_{\vec k}^* v_{\vec k})$,
$\langle S_z\rangle_{\vec k} =\hbar(|u_{\vec k}|^2 -|v_{\vec
k}|^2)$. For circularly polarized light incident onto the surface with
wavevector in the $xz$ plane, $\vec {A}(t)$ = $(A_x \sin \omega t ,
A_y \cos \omega t , A_z \sin\omega t )$ and $\cal{\vec A}$ =
$(-\mathrm{i} A_{x}, A_{y}, -\mathrm{i} A_{z})$. Straightforward
application of time-reversal and crystal symmetries \cite{YHWang} yields
the following expression for the photoemission transition rate:
\begin{eqnarray}\label{m2}
I(\vec{k}) = &a^2& \left( |{\cal A}_x|^2 + |{\cal A}_y|^2 \right)  + b^2 | {\cal A}_z|^2 \nonumber \\
&+& a^2 {\rm Im} \left({\cal A}_x {\cal A}_y^* \right) \langle S_z\rangle_{\vec k} \nonumber \\
 &+& 2 a b  {\rm Im}[{\cal A}_x^* {\cal A}_z\langle S_y \rangle_{\vec k}- {\cal A}_y^* {\cal A}_z\langle S_x \rangle_{\vec k}]
\end{eqnarray}
where $a$ and $b$ are bandstructure dependent complex constants and
$Im$ refers to the imaginary part. Circular dichroism is obtained by
taking the difference of Eq.(\ref{m2}) with opposite photon helicity
($A_y\rightarrow-A_y$):
\begin{equation}\label{diff}
\Delta I = a^2 {\rm Im} ({\cal A}_x {\cal
A}_y^*)\langle S_z \rangle_{\vec k} - 4ab {\rm Im} ({\cal A}_z {\cal
A}_y^*) \langle S_x \rangle_{\vec k}
\end{equation}
Eq.(\ref{diff}) suggests that the modulation in the difference spectra is due to the presence of $\langle S_z\rangle$.

\begin{figure}[htb]
\includegraphics[width=\columnwidth]{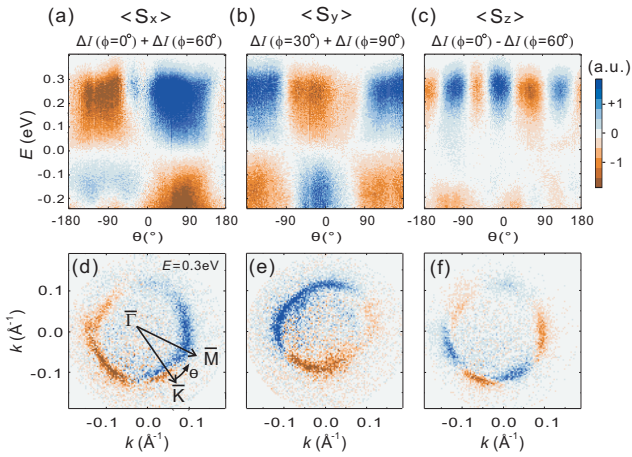}
\caption{Extracted complete spin maps from CD-ARPES on Bi$_2$Se$_3$. (a) $x$, (b) $y$  and (c) $z$ components of the
spin-texture over the complete surface states obtained by summing or
subtracting $\Delta I$ data volumes at different $\phi$.
Color maps were created by integrating the data radially in k-space
over a $\pm0.015 \AA^{-1}$ window about the surface state contours
at each energy. Constant energy cuts at the Fermi level for each
spin component are shown directly below in (d) through (f). From Ref. \cite{YHWang}.}
\label{fig:YHWFig2}
\end{figure}

\subsection{Spin-orbit texture of topological insulators}
The derivation of the CD-ARPES intensity provides a method to extract each spin component.
Although the CD spectra is a linear combination of $\langle S_x \rangle$ and
$\langle S_z\rangle$, they can be disentangled by considering the following symmetries of Bi$_2$Se$_3$(111). Time-reversal symmetry
and three-fold rotational symmetry together dictate that $\langle
S_z \rangle$ flips sign upon a $\Delta\phi=60^{\circ}$ rotation
while $\langle S_x \rangle$ stays unchanged. Taking the difference
(sum) of CD patterns $\Delta\phi=60^{\circ}$ apart isolates $\langle
S_z \rangle$ ($\langle S_x \rangle$). The $\langle S_y \rangle$
component is trivially obtained by performing the procedure for
$\langle S_x\rangle$ under a $90^{\circ}$ sample rotation.

All three spin components are obtained [Fig. \ref{fig:YHWFig2}] by taking the difference or the sum of the difference spectra at various sample rotations [Figs. \ref{fig:YHWFig1}(e)-(h)]. It can be seen from these maps [Figs. \ref{fig:YHWFig2}(a) and (b)] (also see Fig. 3 in Ref. \cite{YHWang}) that at energies within $\pm$ 0.1 eV from the DP, the in-plane spin component follows a sinusoidal dependence on the angle $\theta$ around the constant energy contour. Considering that the out-of-plane component is largely absent in this regime [Figs. \ref{fig:YHWFig2}], the spin vectors form a helical spin texture in the momentum space with the sense of rotation reverses above and below the DP (see Fig. 3 in Ref. \cite{YHWang}). The out-of-plane spin component starts to develop in the high energy regime [Figs. \ref{fig:YHWFig2}(c) and (f)]. The magnitude of the out-of-plane component as a function of energy $\langle S_z \rangle^{0} (E)$ as extracted from Fig. \ref{fig:YHWFig2}(c) agrees well with the prediction of $k \cdot p$ theory to the third order \cite{FuHexWarp}. However, there are also modulations in the in-plane component at the high energy regime [Figs. \ref{fig:YHWFig2}(a),(d)]. Ref. \cite{YHWang} finds that such modulation may be caused by canting of the in-plane spin component (see Fig. 4 in Ref. \cite{YHWang}), which is shown by first-principle calculations to be induced by higher order terms in $k \cdot p$ theory \cite{Basak}.

The expectation value of orbital angular momentum can also be related to the CD-ARPES spectra of topological surface states. Through first-principle calculation, Park et al. found that the sinusoidal form of the radially integrated CD spectra is consistent with a non-zero orbital angular momentum \cite{Park2011,Park2012}. The finite orbital angular momentum is essential for the strong Rashba effect of the system \cite{Park2011}, giving rise to the spin-momentum locking of the surface states. One interesting result of this work is that the orbital angular momentum is found to point in the opposite direction of the spin, which suggests that the surface states of Bi$_2$Se$_3$ have total angular momentum $J=1/2$ \cite{ZhangNatPhy,Park2011}.

\subsection{Generalization to other Rashba systems}

The above treatment of the photoemission matrix element can be generally applied to spin-orbit coupled systems interacting with polarized photons. Therefore, it is equally suitable for the extraction of spin-orbital texture of topologically trivial materials from their CD-ARPES spectra. Among them are the quantum well states and surface states on metal surfaces that exhibit strong Rashba spin-splitting \cite{Bihlmayer}.

\begin{figure*}[htb]
\includegraphics[scale=0.45]{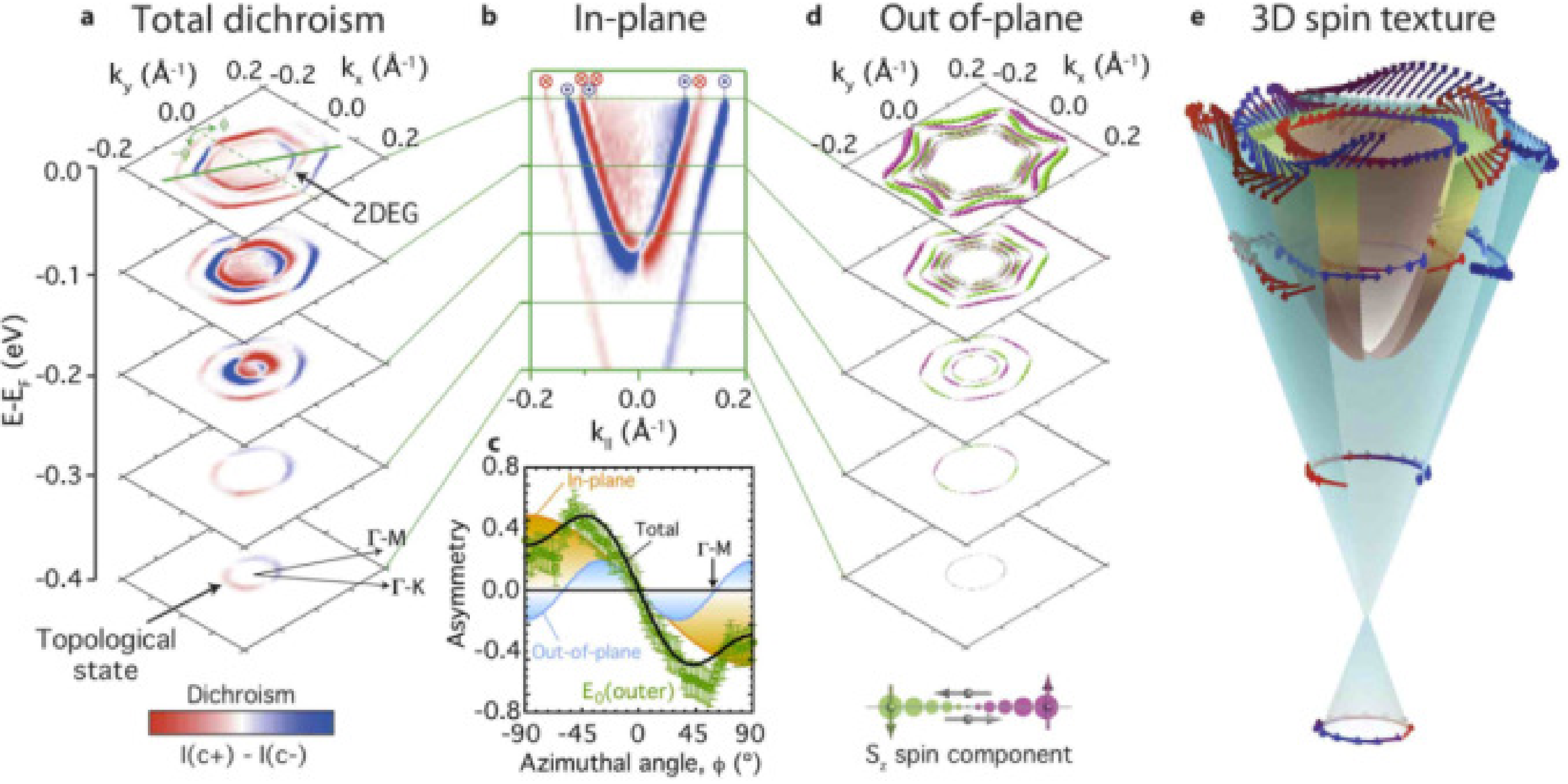}
\caption{Three-dimensional spin texture of the topological and 2DEG states on Bi$_2$Se$_3$. (a,b) Circular dichroism in constant energy contours and E vs. k dispersions measured by ARPES encode the spin texture of the topological and 2DEG states. CD along $\Gamma$-M (b), containing only the in-plane component of the spin texture, revealing opposite helicity for consecutive Fermi surface sheets. (c) Quantitative analysis of the angular ($\phi$) dependence of the dichroism around the Fermi surface, shown for the outer $E_0$ 2DEG state. A large sin(3$\phi$) contribution, also observed for the TSS \cite{Bahramy}, reveals a significant out-of-plane spin canting of the larger Fermi surface sheets, consistent with the calculations shown in (d). In contrast, all other states largely retain the in-plane spin texture characteristic of classic Rashba systems, all the way up to the Fermi energy. Together, this leads to a rich three-dimensional spin-texture of the surface electronic structure of TIs, as summarized for the TSS and lowest ($E_0$) Rashba-split subband of the 2DEG in (e). From Ref. \cite{Bahramy}.} \label{fig:Bahramy}
\end{figure*}

Bahramy et al. \cite{Bahramy} have performed CD-ARPES measurement on both topological surface states and quantum well states. The quantum well states are derived from the two-dimensional electron gas (2DEG) confined in the band-bending region by depositing potassium onto Bi$_2$Se$_3$ \cite{King,Bianchi,Zhu,ParkPRB,HsiehPRL09}. The modulated spin texture of the topological surface states extracted from their CD-ARPES spectra [Fig. \ref{fig:Bahramy}(a)] is consistent with the earlier result \cite{YHWang}. Despite not being topologically protected, the quantum well states also contain rich spin texture with both in- and out-of-plane components [Figs. \ref{fig:Bahramy}(b) and (d)]. Furthermore, the spin texture on consecutive Fermi surface sheets exhibits alternating left-right helicity [Figs. \ref{fig:Bahramy}(b), (d) and (e)]. Such a unique spin texture indicates that the outer and inner branches of the 2DEG states do not merge with each other at large k-vectors, in stark contrast to conventional Rashba systems \cite{Bahramy}.

\begin{figure}[htb]
\includegraphics[width=\columnwidth]{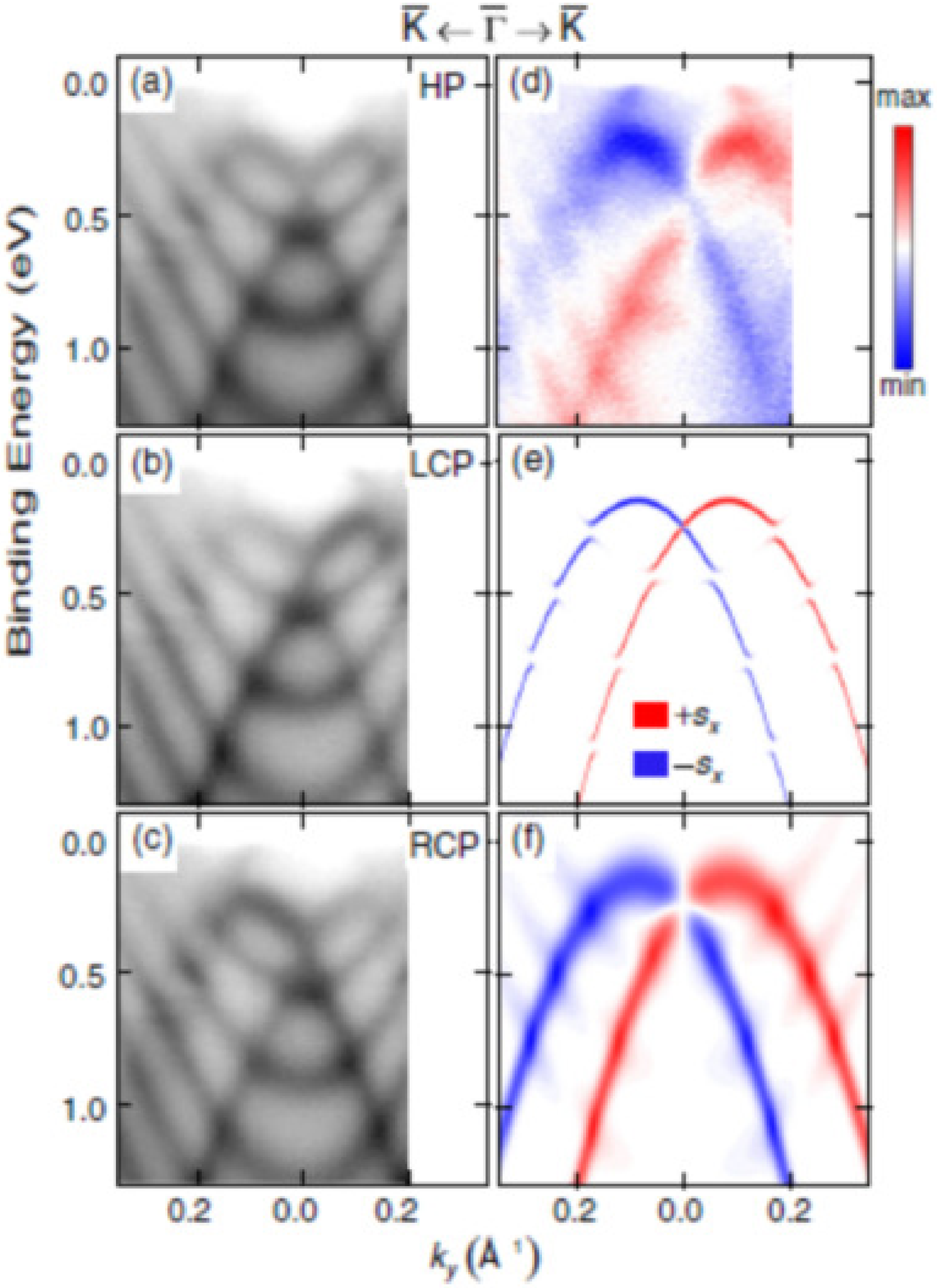}
\caption{Photoemission results taken from Bi/Ag based on a 20 ML Ag film. The data were taken along $K-\Gamma-K$ using 22-eV photons and (a) HP, (b) LCP, and (c) RCP polarization configurations. (d) Dichroic function extracted from the data. (e) Calculated band structure weighted by the spin polarization
inherited from the Rashba split surface states using red and blue colors. (f) Calculated dichroic function. From Ref. \cite{Bian}.} \label{fig:Bian}
\end{figure}

Strong CD-ARPES spectra have been observed on the surface and quantum well states of the thin film heterostructure Bi/Ag(111) \cite{Bian}. Topologically trivial, Bi is a semi-metal with strong Rashba splitting on its surface bands, which can be seen in Fig. \ref{fig:Bian}(a) as the pair of concave parabolic bands. The convex parabolic bands are due to the quantum well states from the 20-ML Ag film. From Figs. \ref{fig:Bian}(d)-(f), we can see that the extracted spin texture of the Bi surface states is consistent with what is measured with spin-resolved ARPES on the same system \cite{Meier,He,Hirahara}. Interestingly, the unpolarized Ag quantum well states also exhibit CD about $\Gamma$ [Fig. \ref{fig:Bian}(d)] which is due to the interference with the surface states \cite{Bian}.

\section{Summary and outlook}

In summary, we have reviewed the recent experiments and theories on using circularly polarized light for ARPES to study the spin-orbit texture of topological materials. These studies have shown that the helicity of light and the experimental geometry directly affects the momentum-dependent surface state photoemission intensity and spin polarization of photoelectrons. The features in the CD-ARPES spectra are quantitatively consistent with matrix element and first principle calculations, which provide a general method to extract vectorial spin-orbit texture from materials with Rashba spin-orbit coupling.

The coupling between the helicity of light and the spin texture of TI suggests many opto-spintronic applications. For example, topological insulator can be exploited as a spin-polarized electron source, from which arbitrary magnitude and direction of spin polarization can be generated by changing the polarization of light \cite{ParkLouie}. There are also numerous proposals to use circularly polarized light to selectively couple to the surface states to generate surface spin waves \cite{Raghu} or spin-polarized currents \cite{Lu,Hosur,McIver}.

\begin{acknowledgement}
The authors would like to thank Dr. David Hsieh for valuable discussions and Dr. Joshua Lui for careful reading of the manuscript. This work is financially supported by Department of Energy Office of Basic Energy Sciences Grant numbers DE-FG02-08ER46521 and DE-SC0006423 (data acquisition and analysis), Army Research Office (ARO-DURIP) Grant number W911NF-09-1-0170 (ARTOF spectrometer) and in part by the Alfred P. Sloan Foundation.

\end{acknowledgement}


\begin{thebibliography}{[1]}

\bibitem{MooreReview} J.\,E. Moore, Nature 464, 194 (2010)
\bibitem{KaneReview} M.\,Z. Hasan, C.\,L. Kane, Rev. Mod. Phys. 82, 3045 (2010)
\bibitem{QiZhangReview} X.-L. Qi, S.-C. Zhang, Physics Today 63, 33 (2010)
\bibitem{Hosur} P. Hosur, Phys. Rev. B 83, 035309 (2011).
\bibitem{HsiehSHG} D. Hsieh, et al., Phys. Rev. Lett. 107, 077401 (2011)
\bibitem{McIver} J.\,W. McIver, et al., Nat. Nano. 7, 96 (2012)
\bibitem{YHWang12} Y.\,H. Wang et al., Phys. Rev. Lett. 109, 127401 (2012)

\bibitem{Hsieh08} D. Hsieh, et al., Nature 452, 970 (2008).
\bibitem{Chen} Y.\,L. Chen, et al., Science (2009).
\bibitem{Xia} Y. Xia, et al., Nature Phys. 5, 398 (2009).

\bibitem{Hufner} S. Hufner, Photoelectron Spectroscopy (Springer, Berlin,2003).

\bibitem{ZhangNatPhy} H. Zhang, et al., Nature Phys. 5, 438 (2009).
\bibitem{HsiehNat09} D. Hsieh, et al., Nature 460, 1101 (2009).

\bibitem{Bychkov} Y.\,A. Bychkov and E.\,I. Rashba, JETP Lett. 39, 78 (1984).
\bibitem{Wolf} S.\,A. Wolf, et al., Science 294, 1488 (2001) 

\bibitem{Souma} S. Souma, et al., Phys. Rev. Lett. 106, 216803 (2011);

\bibitem{Osterwalder} J. Osterwalder, in Magnetism: A Synchrotron Radiation Approach, (Springer, Berlin, 2006).
\bibitem{Meier} F. Meier, et al., Phys. Rev. B 77, 165431 (2008)
\bibitem{Yazyev} O. V. Yazyev, J.\,E. Moore and S.\,G. Louie, Phys. Rev. Lett. 105, 266806 (2010)

\bibitem{Pan} Z.-H. Pan, et al., Phys. Rev. Lett. 106, 257004 (2011).
\bibitem{Xu} S.-Y. Xu, et al., arxiv:1101.3985 (2011)
\bibitem{Jozwiak} C. Jozwiak et al., Phys. Rev. B 84, 165113 (2011).

\bibitem{Sakurai} J.\,J. Sakurai, Modern Quantum Mechanics (Addison-Wesley,1994)
\bibitem{FuHexWarp} L. Fu, Phys. Rev. Lett. 103, 266801 (2009).
\bibitem{KaneMele} C.\,L. Kane and E.\,J. Mele, Phys. Rev. Lett. 95, 146802 (2005).
\bibitem{ParkLouie} C.-H. Park and S. G. Louie, Phys. Rev. Lett. 109, 097601 (2012) 

\bibitem{OpticalOrientation} F. Meier and B. P. Zakharchenya, in Optical Orientation (Modern Problems in Condensed Matter Sciences vol 8). (Elsevier, Amsterdam, 1984)
\bibitem{Westphal} C. Westphal, J. Bansmann, M. Getzlaff, G. Schonhense, Phys. Rev. Lett. 63, 151 (1989);
\bibitem{Vyalikh} D.\,V. Vyalikh, et al., Phys. Rev. Lett. 100, 056402 (2008)
\bibitem{Kaminski} A. Kaminski et al., Nature (London) 416, 610 (2002)
\bibitem{Borisenko} S.\,V. Borisenko, et al., Phys. Rev. Lett. 92, 207001 (2004)
\bibitem{Zabolotnyy} V.\,B. Zabolotnyy, et al., Phys. Rev. B 76, 024502 (2007)
\bibitem{Liu} Y. Liu, et al., Phys. Rev. Lett. 107, 166803 (2012) 

\bibitem{Schneider} C.\,M. Schneider, J. Kirschner, Critical Reviews in Solid State and Materials Sciences. 20, 179 (1995);
\bibitem{Magnetism} J. Stoher, H.\,C. Siegmann, Springer Berlin (2006)
\bibitem{Stoher} J. Stoher, et al., Science 259, 658 (1993).
\bibitem{Schneider91} C.\,M. Schneider, et al., Phys. Rev. B 44, 12066 (1991)

\bibitem{Halilov} S.\,V. Halilov, et al., J. Phys: Condens. Mat. 5, 3851 (1993)
\bibitem{Frentzen} F. Frentzen, et al., Z. Phys. B 100, 575 (1996)
\bibitem{MoorePhotoCurrent} J.\,E. Moore, J. Orenstein, arXiv:0911.3630 (2009)

\bibitem{Mirhosseini} H. Mirhosseini et al., Phys. Rev. Lett. 109, 036803 (2012)
\bibitem{Jung} W. Jung et al., Phys. Rev. B 84, 245435 (2011) 
\bibitem{Park2012} S.\,R. Park et al., Phys. Rev. Lett. 108, 046805 (2012)
\bibitem{Ishida} Y. Ishida et al., Phys. Rev. Lett. 107, 077601 (2011)
\bibitem{Scholz} M.\,R. Scholz et al., arxiv:1108.1053 (2011)
\bibitem{YHWang} Y.\,H. Wang et al., Phys. Rev. Lett. 107, 207602 (2011)

\bibitem{Kuroda} K. Kuroda, et al., Phys. Rev. Lett. 105, 076802 (2010).
\bibitem{Kirchmann} P. Kirchmann et al., Appl. Phys. A 91, S. 211-217 (2008).
\bibitem{Kromker} B. Kromker et al., Rev. Sci. Instrum. 79, 053702 (2008).
\bibitem{Carpene} E. Carpene et al., Rev. Sci. Instrum. 80, 055101 (2009).
\bibitem{Winkelman} A. Winkelmann et al., New J. Phys. 14 043009 (2012).

\bibitem{Bian} G. Bian et al., Phys. Rev. Lett. 108, 186403 (2012)
\bibitem{Dresselhaus} G. Dresselhaus, Phys. Rev. 100, 580 (1955).
\bibitem{Basak} S. Basak, et al., Phys. Rev. B 84, 121401 (2011)

\bibitem{Park2011} S.\,R. Park et al., Phys. Rev. Lett. 107, 156803 (2011)
\bibitem{Bihlmayer} G. Bihlmayer, et al., Surf. Sci. 600, 3888 (2006) 
\bibitem{Bahramy} M.\,S. Bahramy et al., arxiv:1206.0564 (2012) 
\bibitem{King} P.\,D. King et al., Phys. Rev. Lett. 107, 096802 (2011).
\bibitem{Bianchi} M. Bianchi et al., Nat. Comm. 1, 128 (2010).
\bibitem{Zhu} Z.-H. Zhu et al., Phys. Rev. Lett. 107, 186405 (2011).
\bibitem{ParkPRB} S.\,R. Park et al., Phys. Rev. B 81, 041405 (2010).
\bibitem{HsiehPRL09} D. Hsieh, et al., Phys. Rev. Lett. 103, 146401 (2009).

\bibitem{He} K. He et al., Phys. Rev. Lett. 104, 156805 (2010).
\bibitem{Hirahara} T. Hirahara, et al., Phys. Rev. B 76, 153305 (2007).

\bibitem{Raghu} S. Raghu, S.\,B. Chung, X.-L. Qi, S.-C. Zhang, Phys. Rev. Lett. 104, 116401 (2010).
\bibitem{Lu} H.-Z. Lu, W.-Y. Shan, W. Yao, Q. Niu, S.-Q. Shen, Phys. Rev. B 81, 115407 (2010).

\end{thebibliography}
\end{document}